\documentclass[prb,byrevtex,amsmath,amssymb,preprint]{revtex4}

\usepackage{graphicx}
\usepackage{dcolumn}
\usepackage{bm}

\begin{document}
\preprint{PREPRINT}

\title[Short Title]{Adsorption of Oppositely Charged Polyelectrolytes onto a Charged Rod}

\author{Ren\'e Messina}
\email{messina@thphy.uni-duesseldorf.de}
\affiliation
{Institut f\"ur Theoretische Physik II,
Heinrich-Heine-Universit\"at D\"usseldorf,
Universit\"atsstrasse 1,
D-40225 D\"usseldorf,
Germany}

\date{\today}

\begin{abstract}
  The adsorption of highly charged flexible
  polycations  and polyanions on a charged cylindrical substrate is investigated by means of
  Monte Carlo (MC) simulations.
  A detailed structural study, including monomer and fluid charge distributions, is provided.
  The influence of a short
  range attraction between the polycations and the negatively charged substrate is also considered.
  We demonstrate that the building up of multilayer structures is highly prohibited mainly due
  to the high entropy penalty stemming from the low dimensionality of the substrate at strong curvature.
\end{abstract}

\maketitle

\section{Introduction}

The adsorption of charged polymer chains [polyelectrolytes (PEs)]
on oppositely charged surfaces and particles has been the subject
of intense theoretical and experimental research work.
Due to the strong mutual electrostatic attraction, the resulting
complex is highly stable and can exhibit new and interesting  physico-chemical
properties. For instance, such a process is employed to stabilize charged colloidal
suspensions.
Another more elaborated PE adsorption technique, consists of successive
deposition of polycations (PAs) and polyanions (PAs) leading to well controlled PE
multilayer structures. Such materials present enormous technological applications
and this deposition method has been especially exploited with planar \cite{Decher_1997}
and spherical \cite{Caruso_Science_1998} substrates.
In principle this technique could also be applied to (ultra-thin)
\textit{cylindrical substrates}. However due to the very low dimensionality there,
a quantitative experimental characterization of the PE multilayering
is more difficult.

On the theoretical side, there exist a few analytical works about PE multilayers
on charged planar surfaces,\cite{Solis_JCP_1999,Netz_Macromol_1999b,Castelnovo_2000}
but no theory is available for thin cylindrical substrates.
Recent MC simulations were devoted for the PE multilayering on planar
\cite{Messina_Macromol_2003} and spherical \cite{Messina_Langmuir_2003} charged surfaces.
The aim of this paper is to fill this gap and to investigate by means of MC simulations
the structure of PCs and PAs adsorbed onto a charged cylindrical surface. The relevant role 
of the entropy is especially pointed out.

Our paper is organized as follows: Sec.  \ref{ Sec.simu_method}
is devoted to the description of our MC simulation model.
The measured quantities are specified in Sec. \ref{Sec.Target}.
The PC adsorption is studied in Sec. \ref{Sec.monolayer}.
Then the adsorption of both PC and PA chains is addressed in Sec. \ref{Sec.bilayer}.
Finally, Sec. \ref{Sec.conclu} provides brief concluding remarks.

\section{Simulation model
 \label{ Sec.simu_method}}

The setup of the system under consideration is similar to those recently
investigated with spherical \cite{Messina_Langmuir_2003} and planar
\cite{Messina_Macromol_2003} substrates.
Within the framework of the primitive model we consider a PE
solution near a charged hard rod with an implicit solvent (water)
of relative dielectric permittivity $\epsilon_{r}\approx 80$.
This \textit{fixed} charged rod (whose axis is located at $r=\sqrt{x^2+y^2}=0$) is characterized by a
negative  linear bare charge density $-\lambda_0 e$, where $e$ is the
(positive) elementary charge and $\lambda_0>0$ is the number of charges per unit length.
Electroneutrality is always ensured by the presence of explicit monovalent
($Z_c=1$) rod's counterions of diameter $a$.
PE chains ($N_+$ PCs and $N_-$ PAs)
are made up of $N_m$ \textit{monovalent} monomers ($Z_{m}=1$) of diameter $a$.
Hence, all microions are monovalent: $Z=Z_c=Z_m=1$ and have the same diameter $a$.
For the sake of simplicity, we only consider here symmetrical complexes where
PC and PA chains have the same length and carry the same charge in absolute value.
Here the radius $r_{rod}$ of the charged rod is also $a/2$.
To keep further simplicity in our model, we also assume the same linear charge density
for the rod $- \lambda_0 e = - 1e/a$  as the flexible polyelectrolytes. Hence the only
differences between the rod and the PE chains are the chain stiffness and and the chain length,
where both of them are infinite for the rod.
%
%

All these particles making up the system are confined in a $R \times L$ cylindrical box.
Periodic conditions are applied in the $z$ direction, whereas \textit{hard} coaxial
cylinders are present at $r=a/2$ (location of the cylindrical charged wall) and at $r=R$
(location of an \textit{uncharged} cylindrical wall).
To avoid the appearance of image charges,\cite{Fixman_Macromol_1978,Messina_image_2002}
we assume that on both parts of the charged rod (at $r=a/2$)
the dielectric constants are the same.

The total energy of interaction of the system can be written as

\begin{eqnarray}
\label{eq.U_tot}
U_{tot} & = &
\sum_i  \left[ U_{hs}^{(rod)}(r_i) + U_{coul}^{(rod)}(r_i) + U_{vdw}^{(rod)}(r_i) \right] +
\\ \nonumber
&& \sum _{i,i<j} \left[U_{hs}(R_{ij}) + U_{coul}(R_{ij}) + U_{FENE}(R_{ij}) + U_{LJ}(R_{ij}) \right],
\end{eqnarray}
where the first (single) sum stems from the interaction between an ion $i$
(located at $r=r_i$) and the charged plate,
and the second (double) sum stems from the pair interaction between ions $i$ and $j$
with $R_{ij}=|{\bf R}_i - {\bf R}_j|$ where
${\bf R}_i= x_i {\bf e}_x + y_i {\bf e}_y + z_i {\bf e}_z$.
All these contributions to $U_{tot}$ in Eq. (\ref{eq.U_tot})
are described in detail below.

Excluded volume interactions are modeled via a hardcore potential
 defined as follows
%
\begin{equation}
\label{eq.U_hs}
U_{hs}(R_{ij})=\left\{
\begin{array}{ll}
\infty,
& \mathrm{for}~R_{ij} < a \\
0, & \mathrm{for}~R_{ij} \geq a
\end{array}
\right.
\end{equation}
%
for the microion-microion one, except for the monomer-monomer one,\cite{note_HS} and
%
\begin{equation}
\label{eq.U_hs_rod}
U_{hs}^{(rod)}(z_i)=\left\{
\begin{array}{ll}
\infty,
& \mathrm{for} \quad r_i < r_{rod} + a/2\\
\infty,
& \mathrm{for} \quad r_i > R - a/2\\
0,
& \mathrm{for} \quad r_{rod} + a/2 \leq r_i \leq  R - a/2
\end{array}
\right.
\end{equation}
%
for the rod-microion one.
Hence the minimal radius, $r_0$, of closest approach between the rod-axis and
a microion is $r_0=r_{rod}+ a/2 = a$.

The electrostatic energy of interaction between two ions $i$ and $j$ reads
%
\begin{equation}
\label{eq.U_coul}
\frac{U_{coul}(R_{ij})}{k_BT} =
\pm  \frac{l_B}{R_{ij}},
\end{equation}
%
where +(-) applies to charges of the same (opposite) sign, and
$l_{B}=e^{2}/(4\pi \epsilon _{0}\epsilon _{r}k_{B}T)$ is the Bjerrum
length corresponding to the distance at which two monovalent ions
interact with $k_B T$.
The electrostatic energy of interaction between an ion $i$ and the
(uniformly) charged rod reads
%
\begin{equation}
\label{eq.U_coul_rod}
\frac{U_{coul}^{(rod)}(r_i)}{k_BT} =
\pm  2 l_B \lambda_0  \ln (r_i/r_{rod})
\end{equation}
%
where +(-) applies to positively (negatively) charged ions.
An appropriate and efficient modified Lekner sum was utilized to compute
the electrostatic interactions with periodicity in \textit{one}
direction. 
\cite{Brodka_MolPhys_2002b}
To link our simulation parameters to experimental units and room
temperature ($T=298$K) we choose $a =4.25$ \AA\ leading to the
Bjerrum length of water $l_{B}=1.68a =7.14$ \AA\ and to the
so-called Manning condensation parameter
\cite{Manning_JCP_1969,Manning_BBPC_1996} $\xi_M = l_B/a = 1.68$.

The polyelectrolyte chain connectivity is modeled by employing a standard
FENE potential in good solvent (see, e.g., Ref. \cite{Kremer_FENE_1993}),
which reads
%
\begin{equation}
\label{eq.U_fene}
U_{FENE}(R_{ij})=
\left\{ \begin{array}{ll}
\displaystyle -\frac{1}{2}\kappa R^{2}_{0}\ln \left[ 1-\frac{R_{ij}^{2}}{R_{0}^{2}}\right] ,
& \textrm{for} \quad R_{ij} < R_0 \\ \\
\displaystyle \infty ,
& \textrm{for} \quad R_{ij} \geq R_0 \\
\end{array}
\right.
\end{equation}
%
with $\kappa = 27k_{B}T/ a^2$, $R_{0}=1.5 a$, and $i$ and $j$ stand for two consecutive
monomers along a chain.
The excluded volume interaction between chain monomers is taken into
account via a purely repulsive Lennard-Jones (LJ) potential given
by
%
\begin{equation}
\label{eq.LJ}
U_{LJ}(R_{ij})=
\left\{ \begin{array}{ll}
\displaystyle
4\epsilon \left[ \left(\frac{a}{R_{ij}}\right)^{12}
-\left( \frac{a}{R_{ij}}\right) ^{6}\right] +\epsilon,
& \textrm{for} \quad R_{ij} \leq 2^{1/6} a \\ \\
0,
& \textrm{for} \quad  R_{ij} > 2^{1/6} a
\end{array}
\right.
\end{equation}
%
where $\epsilon=k_BT$.
These parameter values lead to an equilibrium bond length $ l=0.98a$.

An important interaction in PE multilayering is the \textit{non}-electrostatic
\textit{short ranged attraction}, $U_{vdw}^{(rod)}$, between the cylindrical macroion and the PC chain.
To include this kind of interaction, we choose without loss of
generality a (microscopic site-site) van der Waals (VDW) potential of interaction between
the cylindrical macroion and a PC monomer that is given by
%
\begin{equation}
\label{eq.U_vdw}
U_{vdw}^{(rod)}(r_i)=-\epsilon \chi_{vdw}
\left( \frac{a}{r_i - r_0 + a} \right)^6
\hspace{0.5cm} \textrm{for}~ r_i \geq r_0,
\end{equation}
%
\begin{table}
\caption{
Model simulation parameters with some fixed values.
}
\label{tab.simu-param}
\begin{ruledtabular}
\begin{tabular}{lc}
 Parameters&
\\
\hline
 $T=298K$&
 room temperature\\
 $-\lambda_0 e = -40e/L = -e/a$ &
 macroion linear charge density\\
 $Z=1$&
 microion valence\\
 $a =4.25$ \AA\ &
 microion diameter\\
 $l = 0.98a$ &
 bond length\\
 $r_{rod} = a/2$ &
 rod radius\\
 $r_0 = r_{rod} + a/2 = a$ &
 radius of closest approach\\
 $l_{B}=1.68a =7.14$ \AA\ &
 Bjerrum length\\
 $R=40 a $&
 $(\sqrt{x^2+y^2})$-box radius\\
 $L=40 a $&
 $z$-box length\\
 $N_+$&
 number of PCs\\
 $N_-$&
 number of PAs\\
 $N_{PE}=N_+ + N_-$&
 total number of PEs\\
 $N_m = 20$&
 number of monomers per chain\\
 $\chi_{vdw}$&
 strength of the specific VDW attraction
\end{tabular}
\end{ruledtabular}
\end{table}
%
where $\chi_{vdw}$ is a positive dimensionless parameter describing the
strength of this attraction. Thereby, at contact (i.e., $r=r_0$),
the magnitude of the attraction is $\chi_{vdw} \epsilon=\chi_{vdw} k_BT$ which
is, in fact, the relevant characteristic of this potential.\cite{note_VDW}
Since it is not straightforward to directly link this strength of
adsorption  to experimental values, we choose $\chi_{vdw}=5$
as  considered previously.\cite{Messina_Langmuir_2003,Messina_Macromol_2003}

All the simulation parameters are gathered in Table \ref{tab.simu-param}.
The set of simulated systems can be found in
Table \ref{tab.simu-runs}.
The equilibrium properties of our model system were obtained by using standard canonical MC
simulations following the Metropolis scheme.\cite{Metropolis_JCP_1953,Allen_book_1987}
Single-particle moves were considered with an acceptance ratio of
$30\%$ for the monomers and $50\%$ for the counterions.
Typically, about $10^5$ to $10^6$ MC steps per
particle were required for equilibration, and no less than $5 \times
10^5$ subsequent MC steps were used to perform measurements.
To improve the computational efficiency, we omitted the presence
of PE counterions when $N_+=N_-$ so that the system is still
globally electroneutral. We have systematically checked for
$N_+=N_-=6$ (system $B$) that the (average) PE configurations
(especially the monomer distribution) are qualitatively the same
of those where PE counterions are explicitly taken into account,
as it should be.

\begin{table}
\caption{
System parameters. The number of counterions (cations and anions) ensuring
the overall electroneutrality of the system is not indicated. The chain length
$N_m=20$ is always the same as well as the parameters of the cylindrical macroion.
}
\label{tab.simu-runs}
\begin{ruledtabular}
\begin{tabular}{lccc}
 System&
 $N_{PE}$&
 $N_+$&
 $N_-$\\
\hline
 $A$&
 $6$&
 $6$&
 $0$\\
 $B$&
 $12$&
 $6$&
 $6$\\
 $C$&
 $24$&
 $12$&
 $12$\\
 $D$&
 $48$&
 $24$&
 $24$\\
\end{tabular}
\end{ruledtabular}
\end{table}

\section{Measured quantities
 \label{Sec.Target}}

We briefly describe the different observables that are going to be measured.
In order to characterize the PE adsorption, we compute the monomer density
$n_{\pm}(r)$ that is normalized as follows

\begin{equation}
\label{eq.n_r}
\int ^{R-a/2}_{r_0} n_\pm(r) 2\pi rL dr = N_\pm N_m
\end{equation}
%
where $+(-)$ applies to PCs (PAs). This quantity is of special interest
to characterize the degree of ordering in the vicinity of the cylindrical macroion surface.

The total number of  accumulated monomers $\bar{N}_{\pm}(r)$ within a distance $r$
from the axis (located at $r=0$) of the cylindrical macroion is given by
%
\begin{equation}
\label{eq.N_r}
\bar{N}_\pm(r) = \int ^{r}_{r_0} n_\pm(r') 2\pi r'L dr'
\end{equation}
%
where $+(-)$ applies to PCs (PAs).
This observable will be addressed in the study of  PC adsorption.
(Sec. \ref{Sec.monolayer}).

Another relevant quantity is the global \textit{net fluid charge}
$\lambda(r)$ which is defined as follows
\begin{equation}
\label{Eq.Qr}
\lambda(r)=\int ^{r}_{r_0} \left[
\tilde{n}_{+}(r')-\tilde{n}_{-}(r')\right] 2\pi r'dr',
\end{equation}
%
where $\tilde{n}_+$ ($\tilde{n}_-$) stand for the density of all the
positive (negative) microions (i.e., monomers and counterions).
Thus, $\lambda(r)$ corresponds to the (reduced)
net fluid charge per unit length (omitting the bare macroion linear-charge $-\lambda_0$)
within a distance $r$ from the axis of the charged cylindrical wall. At the uncharged
wall, electroneutrality imposes $\lambda(r=R-a/2)=\lambda_0$.
By simple application of the Gauss' law,
$\left[ \lambda(r)-\lambda_0\right]/r$ is directly proportional
to the mean electric field at $r$.  Therefore $\lambda(r)$ can
measure the \textit{screening} strength of the cylindrical macroion-charge
by the neighboring solute charged species.

\section{Adsorption of polycations
 \label{Sec.monolayer}}

In this part, we study the adsorption of PC chains (system $A$)
for the two couplings $\chi_{vdw}=0$ and $\chi_{vdw}=5$. The case
$\chi_{vdw}=0$, corresponding to a purely electrostatic regime, is
well understood for "normal spherical" electrolyte ions  near a
charged rod.
\cite{Manning_JCP_1969,Belloni_CollSurf_1998,Deserno_Macromol_2000}
Nevertheless, the situation is much more delicate for PE
adsorption, and that has not been the subject of numerical
simulations.\cite{note_dias} This is a decisive step to elucidate
the even more complex PE multilayer structures where additionally
PAs are also present.

Here, where $N_-=0$ (i.e., no polyanions), global electroneutrality is ensured
by the presence of explicit PC's counterions (i.e., monovalent anions) and
the macroion-rod's counterions (i.e., monovalent cations).

\begin{figure}
\includegraphics[width = 8.0 cm]{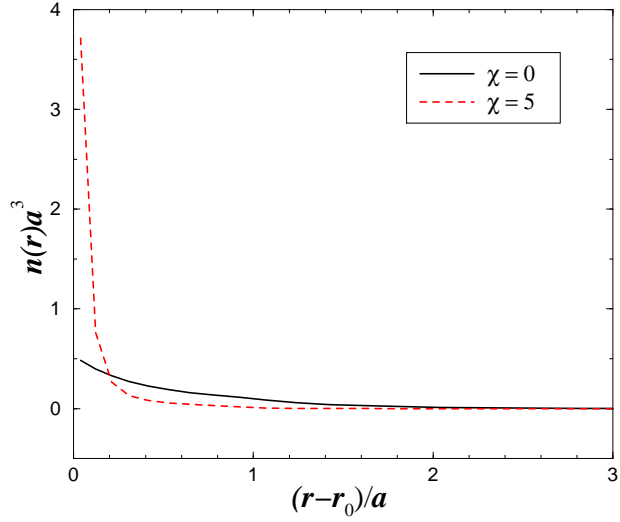}
\caption{
Profiles of PC monomer-density $n_+(r)$ at different $\chi_{vdw}$ couplings (system $A$).
}
\label{fig.nr_monolayer}
\end{figure}
%

\begin{figure}
\includegraphics[width = 8.0 cm]{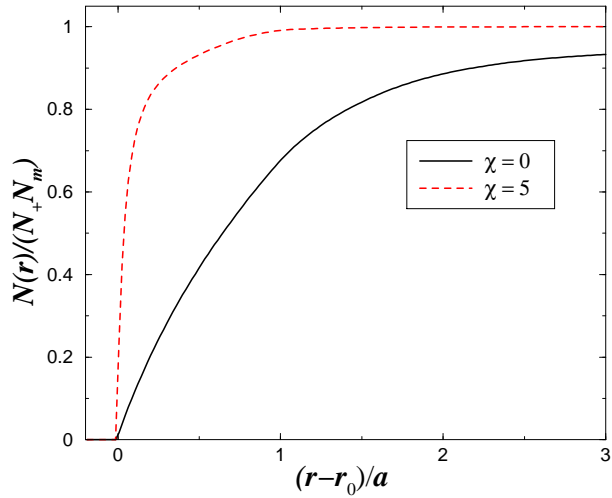}
\caption{
Fraction $\bar{N}_+(z)/(N_+N_m)$ of adsorbed PC monomers at 
different $\chi_{vdw}$ couplings (system $A$).}
\label{fig.Nr_monolayer}
\end{figure}

The profiles of the monomer density $n_+(r)$ are depicted in Fig. \ref{fig.nr_monolayer}.
Near contact (i.e., $r - r_0 \sim 0$),
the density $n_+(r)$ at $\chi_{vdw}=5$ is about one order of magnitude larger than at
$\chi_{vdw}=0$.
To further estimate the degree of adsorption of PC monomers we have plotted
the fraction $\bar{N}_+(z)/(N_+ N_m)$  of adsorbed monomers in Fig. \ref{fig.Nr_monolayer}.
At a radial distance $r-r_0=a$
(corresponding to a width of two monomers), almost all the the monomers (more than 99\%) are
adsorbed for $\chi_{vdw}=5$ against only $\sim 68\%$ for $\chi_{vdw}=0$.
This strong observed PC adsorption at $\chi_{vdw}=0$ is of course due to the high valence of the
polycation (here $N_m=20$) and the intense attractive electric field near the \textit{thin} rod.
On the other hand, at high enough $\chi_{vdw}$ and (even) \textit{without} rod-monomer electrostatic 
interaction [i.e., for a neutral rod ($\lambda_0=0$)], 
the PC adsorption would be very similar to that obtained at finite $\lambda_0$.


\begin{figure}
\includegraphics[width = 8.0 cm]{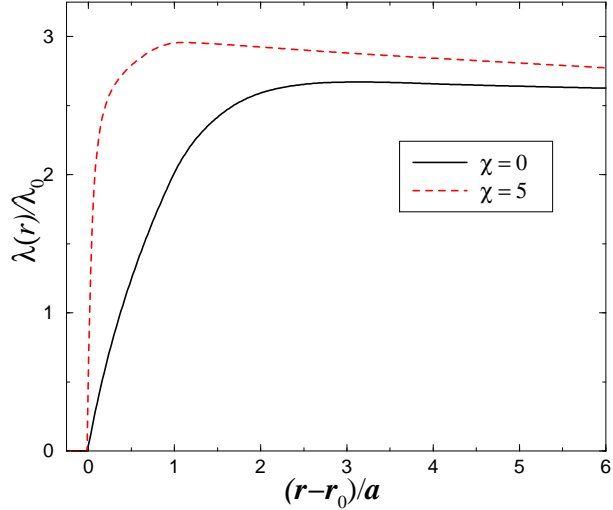}
\caption{ Net fluid charge $\lambda(r)$ at different $\chi_{vdw}$ couplings (system $A$).}
\label{fig.Qr_monolayer}
\end{figure}
%
The (global) net fluid charge $\lambda(r)$ is reported in Fig.
\ref{fig.Qr_monolayer}. In all cases we detect a macroion-surface
charge reversal (i.e., $\lambda(r)/\lambda_0>1$). The position
$r=r^*$ at which $\lambda(r^*)$ gets its maximal value decreases
with $\chi_{vdw}$, due to the $\chi_{vdw}$-enhanced adsorption of
the PCs.  Concomitantly, this \textit{overcharging} increases with
$\chi_{vdw}$, since the (extra) gain in energy by macroion-monomer
VDW interactions can better overcome (the higher $\chi_{vdw}$) the
cost of the self-energy stemming from the adsorbed excess charge.
\cite{Messina_Langmuir_2003,Messina_Macromol_2003} 
More quantitatively, we have $\lambda(r^*-r_0 \approx 1.1a)/\lambda_0
\approx 2.96$ at  $\chi_{vdw}=5$ against $\lambda(r^*-r_0 \approx
3.4a)/\lambda_0 \approx 2.67$ at $\chi_{vdw}=0$. Note that the
maximal value of charge reversal of $(6 \times 20 - 40)/40=200\%$
(i.e., $\lambda(r^*)/\lambda_0 = 3$) allowed by the total charge
of PCs can not be exactly reached due to a slight accumulation of
microanions. However at $\chi_{vdw}=5$, the presence of
microanions near the rod is very marginal, leading in practice to
an overcharging that is fully governed by the PCs. Again this
giant overcharging found at any $\chi_{vdw}$ is due to the high
(intrinsic) electric field generated by the macroion rod. Indeed,
for the same chain parameter, it was found in planar geometry
(with a surface charge density $-\sigma_0e=-0.165 \text{C/m}^2$)
that only a charge reversal of about 70\% [i. e., $\sigma(r^*)/\sigma_0=1.7$] 
could be reached at $\chi_{vdw}=5$.\cite{Messina_Macromol_2003}
For cylindrical substrates, it turns out that the
equivalent surface charge density [here $-\lambda_0e \times L/(2
\pi r_{rod} L) \approx -0.282  \text{C/m}^2$] can be much higher in
experimental conditions. Those (locally) overcharged states should
be the driving force for the building of subsequent PE bilayers
when PA chains are added.
%
\begin{figure*}
\includegraphics[width = 7.0 cm]{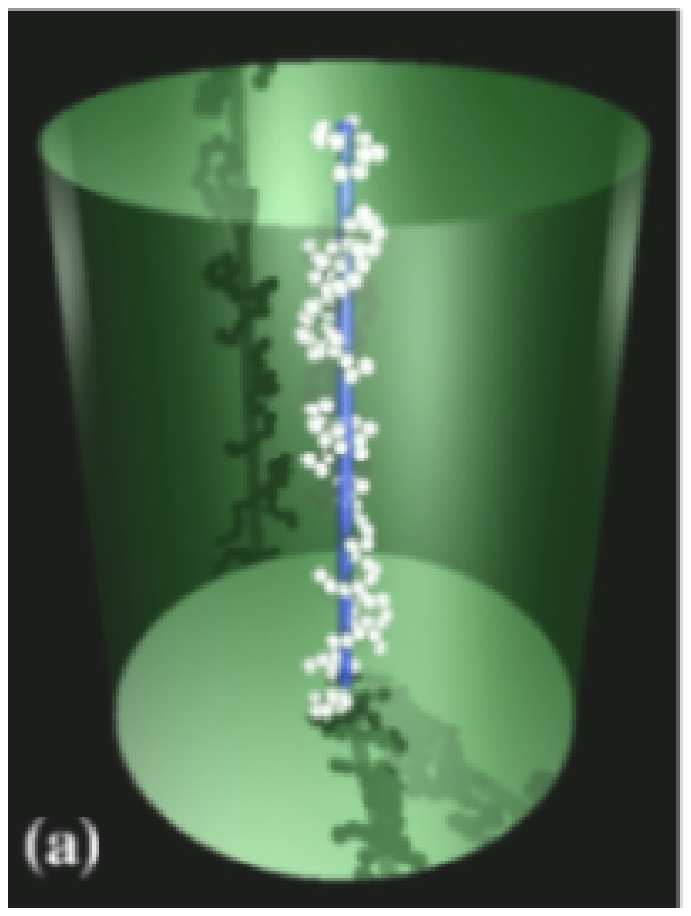}
\includegraphics[width = 7.0 cm]{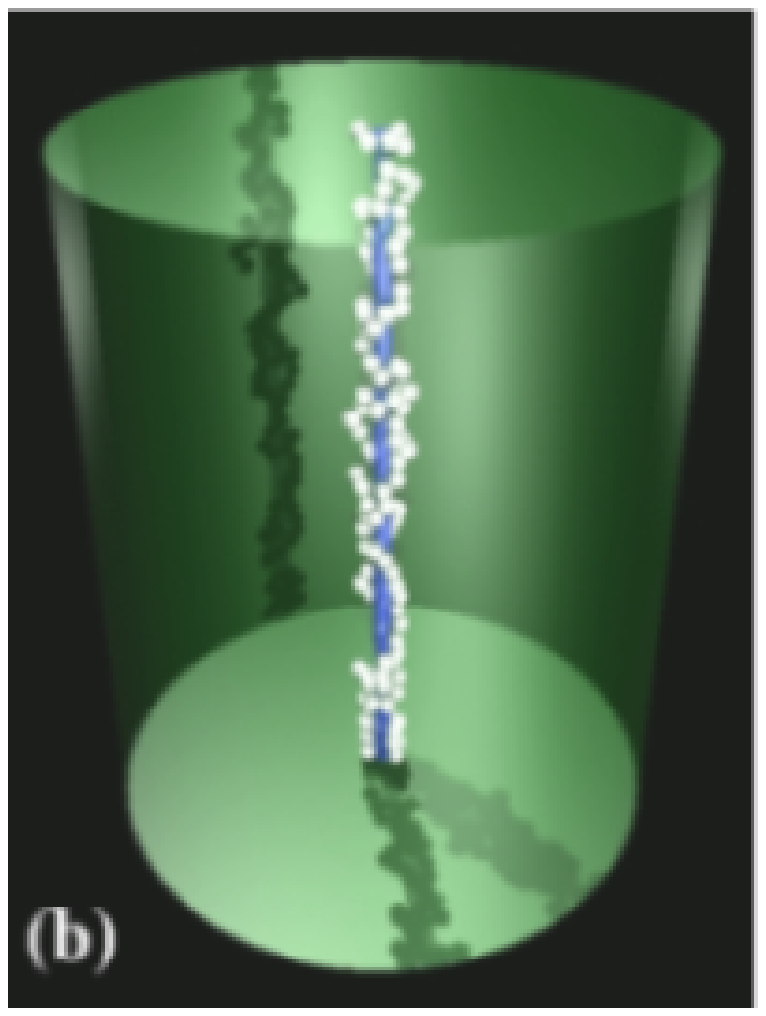}
\caption{
Typical equilibrium configurations for PC chains adsorbed onto an oppositely
charged cylindrical macroion (system $A$).
The little counterions are omitted for clarity.
The outer green cylinder is a guide for the eye.
(a) $\chi_{vdw}=0$
(b) $\chi_{vdw}=5$.
}
\label{fig.snap_monolayer}
\end{figure*}

Typical equilibrium configurations can be seen in Fig.
\ref{fig.snap_monolayer}. The qualitative difference between
$\chi_{vdw}=0$ [Fig. \ref{fig.snap_monolayer}(a)] and
$\chi_{vdw}=5$  [Fig. \ref{fig.snap_monolayer}(b)] is remarkable.
Without additional VDW attraction ($\chi_{vdw}=0$) the adsorption
is more diffuse than at $\chi_{vdw}=5$, where in the latter
situation the $r$-fluctuation is very weak. Interestingly, one can
observe that the PC chains tend to adopt a helical structure. This
kind of conformation is the non-trivial result of two antagonistic
forces: (i) electrostatic correlations and (ii) entropy.
To clarify this point, one has to consider two typical length scales:
\begin{itemize}
\item If the cylindrical macroion is larger or about the PE chain
      (i.e., $L \gtrsim N_ml$ as it is presently the case), then the ground state \cite{note_helix} 
      corresponds to \textit{straight} PE chains sticked to the surface of the rod. 
      At \textit{finite} temperature one obtains PC chains
      spiraling around as the result of the \textit{coupled effects} of entropy and
      electrostatic correlations.
\item If $L \ll N_ml$, then the ground state (already) corresponds to perfectly
      ordered helical structures (so as to overscreen the rod) which locally resemble
      a one-dimensional Wigner crystal. \cite{Nguyen_JCP_2000}
      At finite temperature and strong Coulomb coupling
      (as it is the case with highly charged PEs and macroion-substrates), this high helical
      ordering persists at shorter range.
\end{itemize}
Those reasonings above should at least hold when the equivalent surface charge densities of the PE
and that of the rod are similar, as it is the case in this study.
%
%
\section{Adsorption of polycations and polyanions
 \label{Sec.bilayer}}

We now consider the systems $B-D$ where additionally PA chains are
present, so that we have a (globally) neutral polyelectrolyte
complex (i.e., $N_+=N_-=N_{PE}/2$). Global electroneutrality is
ensured by the counterions of the cylindrical macroion as usual.
We stress the fact that this process is fully reversible for the
parameters under investigation. In particular, we checked that the
same final \textit{equilibrium} configuration is obtained either
by (i) starting from system with isolated PCs and then adding PAs
or (ii) starting directly with the oppositely charged PEs.

\subsection{Monomer distribution
 \label{Sec.monomer}}

\begin{figure}
\includegraphics[width = 8.0 cm]{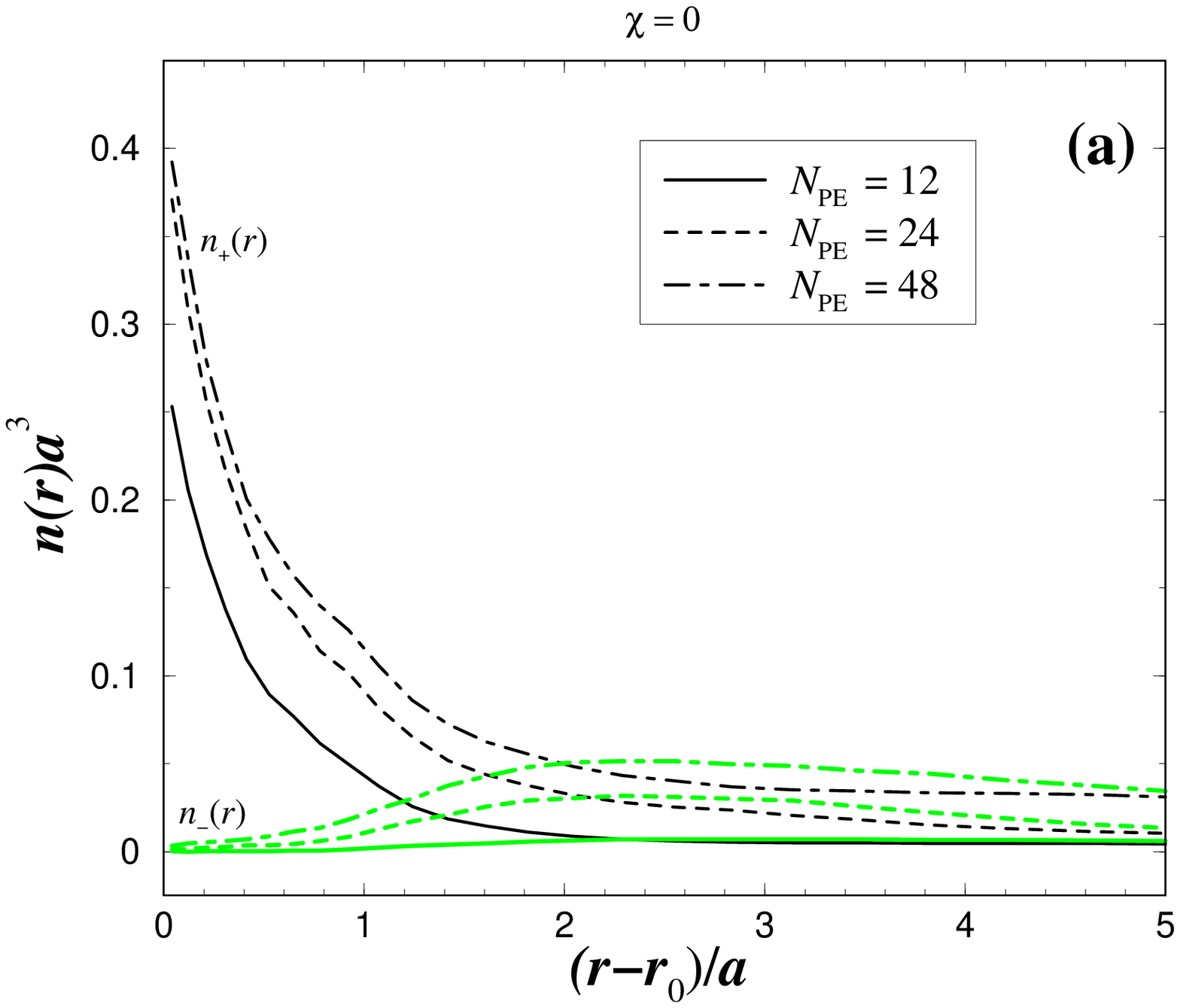}
\includegraphics[width = 8.0 cm]{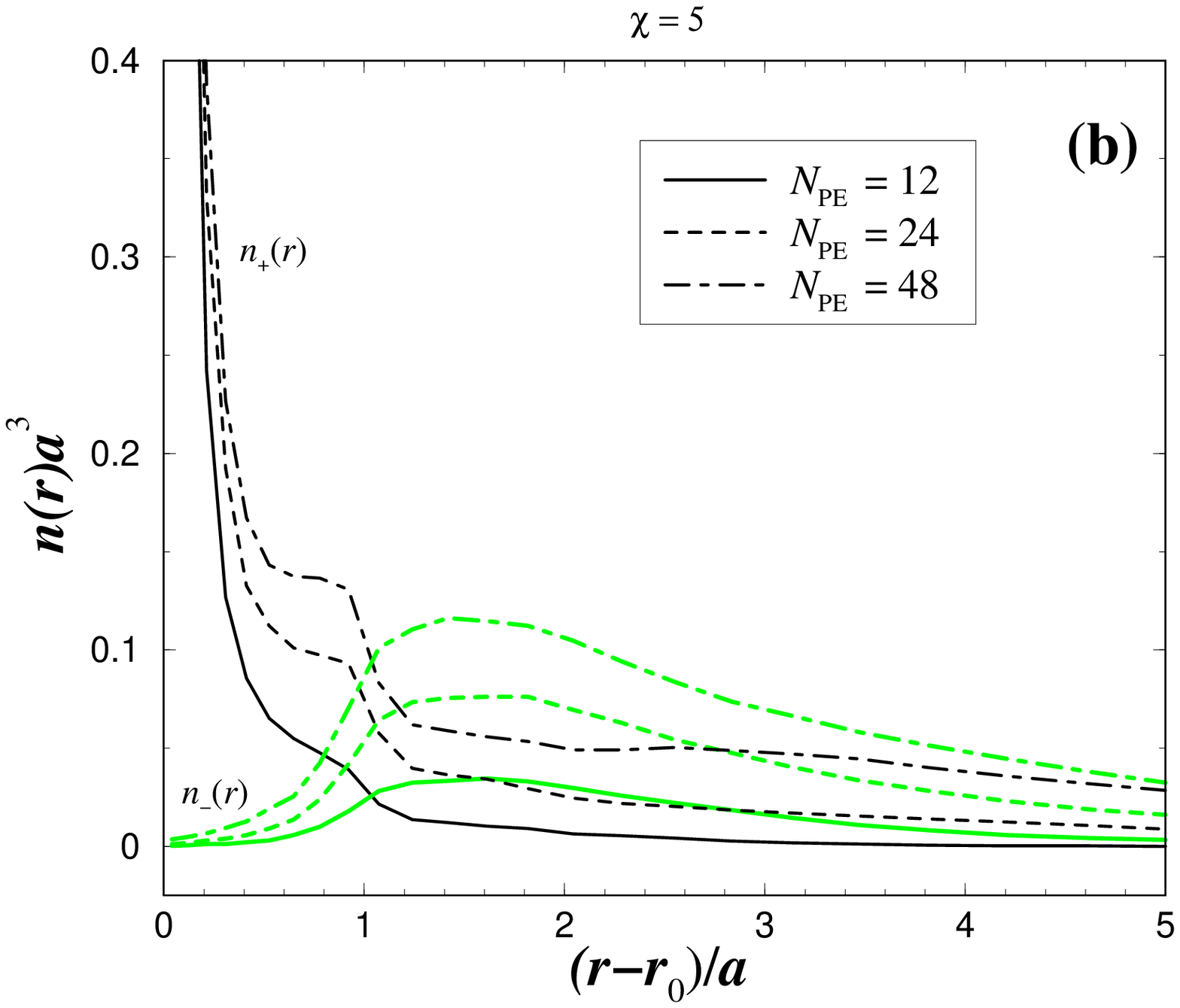}
\caption{
Profiles of monomer density $n_{\pm}(r)$ for oppositely charged polyelectrolytes (systems $B-D$).
(a) $\chi_{vdw}=0$.
(b) $\chi_{vdw}=5$.
}
\label{fig.nr_bilayer}
\end{figure}
%
\begin{figure*}
\includegraphics[width = 13.0 cm]{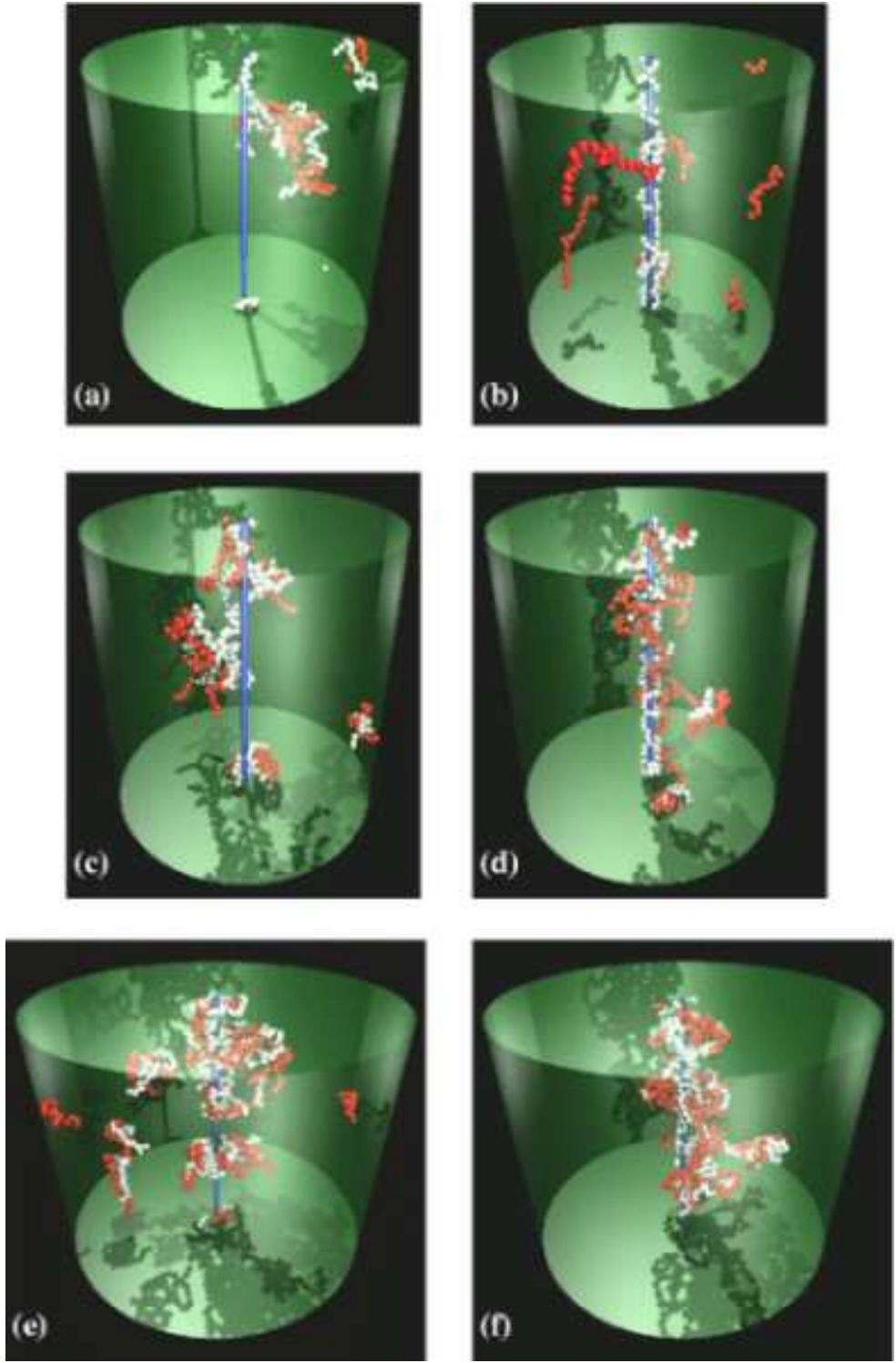}
\caption{
Typical equilibrium configurations for the adsorption of oppositely
charged PE chains (systems $B-D$) onto a cylindrical macroion.
The polycations are in white and the polyanions in red.
The little ions are omitted for clarity.
The outer green cylinder is a guide for the eye.
(a) $\chi_{vdw}=0$, $N_{PE}=12$ (system $B$)
(b) $\chi_{vdw}=5$, $N_{PE}=12$ (system $B$)
(c) $\chi_{vdw}=0$, $N_{PE}=24$ (system $C$)
(d) $\chi_{vdw}=5$, $N_{PE}=24$ (system $C$)
(e) $\chi_{vdw}=0$, $N_{PE}=48$ (system $D$)
(f) $\chi_{vdw}=5$, $N_{PE}=48$ (system $D$).
}
\label{fig.snap_bilayer}
\end{figure*}
%

The profiles of the monomer density $n_\pm(r)$ at $\chi_{vdw}=0$ and $\chi_{vdw}=5$ are
depicted in Fig. \ref{fig.nr_bilayer}(a) and Fig. \ref{fig.nr_bilayer}(b), respectively.
The corresponding microstructures are sketched in Fig. \ref{fig.snap_bilayer}.

\subsubsection{Case of zero-$\chi_{vdw}$}

At $\chi_{vdw}=0$, a comparison between systems $A$ (see Fig.
\ref{fig.nr_monolayer}) and $B$ indicates that the adsorption of
PC monomers is weaker when additional PAs are present. This effect
was already observed with spherical  \cite{Messina_Langmuir_2003}
and planar \cite{Messina_Macromol_2003} substrates, and the same
mechanisms apply here to cylindrical substrates. More explicitly,
the PC  tends to build up a globular state (reminiscent of the
classical \textit{bulk} PE collapse) by getting complexed to the
PA chain, as perfectly illustrated in Fig.
\ref{fig.snap_bilayer}(a). Thereby, the mean number of nearest
monomer-neighbors gets higher which is \textit{both} (i)
entropically and (ii) energetically (at least from the PE complex
viewpoint) favorable. It turns out that for macroion-rods with
large curvature, this PC desorption is counter-intuitively much
higher than in the planar case, \cite{Messina_Macromol_2003}
although a much weaker charge reversal (without PAs) was found for
the latter. Indeed, we just saw that the overcharging by PCs alone
can be quite huge (of the order of 200 \% even at $\chi_{vdw}=0$ -
see Fig. \ref{fig.Qr_monolayer}), and thus, one would naively
expect a new wave of charge reversal in system $B$ (where
$N_+=N_-=6$) of the same order (see also Sec. \ref{Sec.charge} for
a more quantitative description) which in turn should lead to a
(very) stable bilayering. But Figs. \ref{fig.nr_bilayer}(a) and
\ref{fig.snap_bilayer}(a) demonstrate that it is not the case.
Furthermore, an inspection of the density profile $n_-(r)$ of the
PA monomers reveals that there is \textit{no} peak for
$N_{PE}=12$.
Those interesting findings allow us to already draw a partial
conclusion:
\begin{itemize}
\item At large enough rod-curvature and small enough PE concentration, even if a giant overcharging
      would occur in the absence of PAs, it is not sufficient to promote bilayering.
\end{itemize}
Hence, this spectacular PA-induced PC desorption is due to
entropic effects. More precisely, the adsorption onto quasi
one-dimensional objects is extremely costly from an entropy
viewpoint, and if the number of available dipoles (i.e., PC-PA ion
pairs) is too small then ion-dipole correlations are not strong
enough to overcome the ``entropic barrier".

Upon increasing $N_{PE}$ (i.e., the PE concentration) one sees that the density $n_+(r)$
at contact increases [also identifiable on Figs.~\ref{fig.snap_bilayer}(a), (c) and (e)] and
gets gradually less sensitive to $N_{PE}$, showing that a "steady" regime should be reached at 
large enough $N_{PE}$.
Concomitantly, by increasing $N_{PE}$, Fig. \ref{fig.nr_bilayer}(a) shows that
the adsorption of PA monomers increases
[also identifiable on Figs.~\ref{fig.snap_bilayer}(a), (c) and (e)].
This $N_{PE}$-enhanced adsorption of monomers is essentially governed by two concomitant mechanisms:
\begin{itemize}
\item At higher $N_{PE}$, the global ion-dipole correlation gets enhanced 
      (due to an increase of PC-PA ion pair concentration) which favors 
      a stronger monomer adsorption.
\item Concomitantly, at higher $N_{PE}$, where  the available volume becomes lower,
      the \textit{entropy loss} undergone upon adsorption gets reduced, which also favors 
      a stronger monomer adsorption.
\end{itemize}
Nevertheless, only  a very broadened peak in
$n_-(r)$ (at $N_{PE}=48$) is seen and \textit{no second} peak in $n_+(r)$ appears,
indicating a weakly stable bilayer. This feature strongly
contrasts with our previous findings for planar substrates \cite{Messina_Macromol_2003} 
where \textit{stable} trilayers were reported at $\chi_{vdw}=0$ (also
with $N_m=20$). Therefore, we can state:
\begin{itemize}
\item PE multilayering in the electrostatic regime is highly inhibited at large
      rod-curvature, mainly due to entropic effects.
      It is to say that the (conformational) PE entropy along the rod is much smaller than that in
      the bulk and also smaller than that on a planar surface.
\end{itemize}

\subsubsection{Case of finite $\chi_{vdw}$}

In general all our mechanisms previously described at
$\chi_{vdw}=0$ are also present here at  $\chi_{vdw}=5$, but one
has now to take into account the possible dominance of the VDW
interaction near the surface of the cylindrical macroion.

At $\chi_{vdw}=5$, Fig. \ref{fig.nr_bilayer}(b) shows that the
first PC layer is always very stable (despite of the presence of
PAs) due to the the sufficiently strong rod-PC VDW attraction. In
parallel, $n_+(r)$ and in general the ordering increase with
$N_{PE}$, as it should be. At large $N_{PE}=48$, one can detect a
second broadened peak in $n_+(r)$ which could be the signature of
the onset of a third PE layer. However, the height of this peak is
still significantly smaller than the corresponding decaying PA
density $n_-(r)$ in this region. A visual inspection of Fig.
\ref{fig.snap_bilayer}(b), (d) and (f) confirms all those
features. Hence, even with the help of a specific VDW attractive
interaction, it seems that  the formation of stable
structures beyond bilayers is highly inhibited at strong
confinement. This again proves how important is the role of entropy
in PE multilayering with low dimensional substrates.

\subsection{Charge distribution
 \label{Sec.charge}}

\begin{figure}
\includegraphics[width = 8.0 cm]{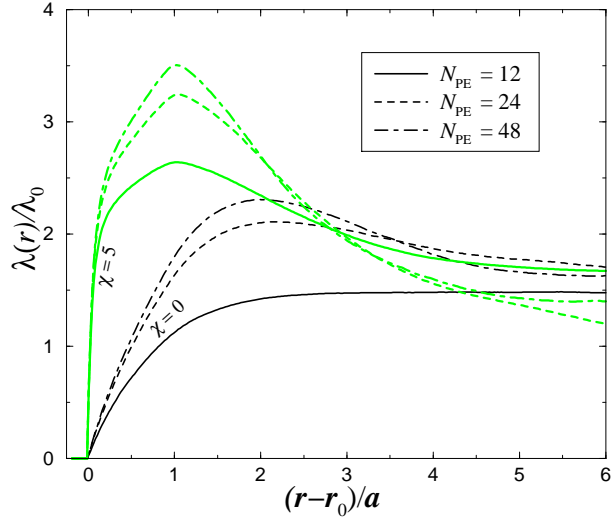}
\caption{ Net fluid charge $\lambda(r)$ at different $\chi_{vdw}$ couplings (systems $B-D$).}
\label{fig.Qr_bilayer}
\end{figure}

The net fluid charge $\lambda(r)$ is reported in Fig. \ref{fig.Qr_bilayer}.
In all cases, the cylindrical macroion gets (locally) overcharged.
Surprisingly, beyond the PC overcharging layer the fluid charge
does not reenter below unity (at least for $r<6a$), proving that
there are no charge \textit{oscillations} in the vicinity of the
rigid rod. Even with an overcharging of
$\lambda(r^*)/\lambda_0=3.5$ for $\chi_{vdw}=5$ and $N_{PE}=48$,
no charge oscillation is generated. This is, nevertheless,
consistent with our previous results of Fig. \ref{fig.nr_bilayer}
where no multilayering was observed. This phenomenon is due to the
fact that our PE concentration is still (very) small even at
$N_{PE}=48$, which is also the experimental situation in the
process of PE multilayering where a \textit{rinsing} is applied at
each successive layer deposition \cite{Decher_1997}. It is well
known that at \textit{high finite} electrolyte concentration,
charge oscillations take place near a charged cylindrical wall.
\cite{Marcelo_JCP_1985,Deserno_Macromol_2000,Deserno_JBCB_2001}

Our results clearly show that the degree of overcharging is
systematically larger at high $\chi_{vdw}=5$, as also observed without PAs
(see Fig. \ref{fig.Qr_monolayer}).
The values of $r^*$ are also systematically shifted to smaller ones as one increases
$\chi_{vdw}$.
However, the position of the peak
($r^*-r_0 \approx a$) for $\chi_{vdw}=5$ remains nearly independent of $N_{PE}$ in contrast
to what happened at $\chi_{vdw}=0$, where $r^*$ is systematically shifted to smaller radial distances
upon increasing $N_{PE}$.
All those observations are consistent with our mechanisms presented so far. 

Nonetheless, as soon as \textit{oppositely charged polyions} can
interact, there is a subtle interplay between clustering and the
lateral correlations of polyions that governs the degree of
overcharging near the macroion surface, as it was already
demonstrated in Refs.
\cite{Messina_Langmuir_2003,Messina_Macromol_2003}. Additionally,
at strong rod curvature, entropic effects become also
considerable. 
Consequently, the physics of overcharging involved in the adsorption of oppositely charged 
PEs onto one-dimensional substrates is highly complicated.


\section{Concluding remarks
 \label{Sec.conclu}}
In summary, we have carried out MC simulations to elucidate the adsorption
of flexible highly charged polycations and polyanions onto a charged cylindrical substrate.
Within the dilute regime, we considered the influence of PE concentration.
In order to enhance the possible formation of PE multilayers we have also considered
an extra non-electrostatic short-range attraction
(characterized here by $\chi_{vdw}$) between the cylindrical macroion surface and the polycations.

As far as the PC adsorption is concerned (in the absence of PAs), we demonstrated that
huge macroion charge reversal occurs even in a purely electrostatic regime with $\chi_{vdw}=0$.
By adding exactly the same amount of PAs, we surprisingly observe a (relatively) marginal
overcharging which is due to (i) PC-PA clustering and (ii) above all to \textit{entropic} effects.

At higher number of PEs, our results show that true bilayering
(i.e.; flat and dense PE layers) can only occur at finite
$\chi_{vdw}$, in contrast to what was recently found with planar
substrates. \cite{Messina_Macromol_2003}
Even at finite
$\chi_{vdw}$, our simulation data demonstrate that stable
multilayering (beyond bilayering) is hard to reach at large
macroion rod-curvature, due to the \textit{high entropy loss}
there. This latter in turn inhibits the appearance of charge
\textit{oscillations}.

However, there must be a certain regime of curvature, typically
when the rod radius is larger or about the chain size (i.e.,
$r_{rod} \gtrsim{N_ml}$), where such entropy effects become less
relevant and would then lead to similar behaviors to those
occurring with planar substrates. In this respect, a future work could include the
effect of rod-curvature on the formation of PE multilayers.
We hope that this contribution will
trigger new theoretical as well as experimental investigations.


\acknowledgments The author thanks C. N. Likos and H. L\"owen for helpful 
discussions and a critical reading of the manuscript.


\begin{thebibliography}{10}
\bibitem{note_HS}
  Only the monomer-monomer excluded volume interaction was not modeled by a
  hard-sphere potential. There, a purely repulsive truncated and shifted 
  Lennard-Jones potential was used [see Eq.\eqref{eq.LJ}] so as to be compatible 
  with the FENE potential employed to generate the chain connectivity.
\bibitem{note_helix}
In this reasoning, we omit the presence of the little counterions which would
  lead to a spurious collapse at zero temperature.
\end{thebibliography}

\end{document}